\documentstyle[aps,prd,psfig,preprint]{revtex}
\begin{document}
\draft

\title{Vacuum defects without a vacuum}
\author{Inyong Cho\footnote[1]{Electronic address: cho@cosmos2.phy.tufts.edu}
and Alexander Vilenkin\footnote[2]{Electronic address: vilenkin@cosmos2.phy.tufts.edu}}

\address{Institute of Cosmology,
        Department of Physics and Astronomy,\\
        Tufts University,
        Medford, Massachusetts 02155, USA}
\date{\today}
\maketitle

\begin{abstract}
Topological defects can arise in symmetry breaking models
where the scalar field potential $V(\phi)$ has no minima 
and is a monotonically decreasing function of $|\phi|$.
The properties of such vacuumless defects are quite different
from those of the ``usual'' strings and monopoles.
In some models such defects can serve as seeds for structure
formation, or produce an appreciable density of
mini-black holes.
\end{abstract}


Symmetry-breaking phase transitions in the early universe can result in the
formation of topological defects: domain walls, strings, monopoles,
and their hybrids~\cite{Kibble,Vilenkin}.
A prototypical symmetry breaking model is of the form

\begin{equation}
{\cal L} = {1 \over 2}\partial_\mu\phi_a\partial^\mu\phi_a - V(f)\;,
\label{eq=lag}
\end{equation}
where $\phi_a$ is a set of scalar fields, $a=1,...,N$, 
$f=(\phi_a\phi_a)^{1/2}$, and the potential $V(f)$ has a minimum
at a non-zero value of $f$.
Domain walls, for example, are obtained for a singlet scalar field, $N=1$.
Domain wall solutions, $\phi =\phi (x)$, interpolate between
$-\eta$ at $x \to -\infty$ and $+\eta$ at $x\to +\infty$,
where $f=\eta$ is the minimum of $V(f)$.

In this paper, we are going to consider defects in models where
$V(f)$ has a local maximum at $f=0$ but no minima;
instead, it monotonically decreases to zero at $f\to\infty$.
Potentials with this asymptotic behavior can arise, for example, 
due to non-perturbative effects in supersymmetric gauge 
theories~\cite{Seiberg}.
In a cosmological context, they have been discussed in the so-called
quintessence models~\cite{Peebles}. Defects arising in such models
could be called ``quintessential'', but we shall resist this temptation 
and use a more descriptive term ``vacuumless''.
We shall see that the properties and evolution of vacuumless defects can be 
quite different from the usual ones.

We begin with the simplest case of a domain wall described by a real scalar
field $\phi$ with a potential

\begin{equation}
V(f) = \lambda M^4 \cosh^{-2}(f/M)\;.
\label{eq=Vwallcosh}
\end{equation}
The field equation 

\begin{equation}
f''(x) = V'(f)
\end{equation}
has a domain wall solution 

\begin{equation}
\sinh(\phi/M) = (2\lambda)^{1/2}Mx
\end{equation}
with a finite energy per unit area

\begin{equation}
\sigma = \int T_0^0 dx = \pi (2\lambda)^{1/2} M^3\;.
\end{equation}
Similarly, for an inverse power-law potential

\begin{equation}
V(f) = \lambda M^{4+n} ( M^n + f^n)^{-1}
\label{eq=Vwallpower}
\end{equation}
one finds $f \propto |x|^{2/(n+2)}$ and $T_0^0(x) \propto |x|^{-2n/(n+2)}$
at $|x| \to \infty$.
For $n>2$, this energy distribution is integrable and the mass per unit
area $\sigma$ is finite.
In this respect, vacuumless walls are quite similar to ordinary
domain walls.

The situation  with strings and monopoles is more interesting.
Let us first consider global strings described by a scalar doublet
$(\phi_1,\phi_2)$ with a power-law potential~(\ref{eq=Vwallpower}).
In the cylindrical coordinates $(r,\theta,z)$, the string ansatz is
$\phi_1 = f(r)\cos\theta$, $\phi_2 = f(r)\sin\theta$,
and $f(r)$ satisfies the equation

\begin{equation}
f'' + {f' \over r} -{f \over r^2} - {dV \over df} =0\;.
\label{eq=feqgb}
\end{equation}
The asymptotic solution of Eq.~(\ref{eq=feqgb}) at 
$r\gg \lambda^{-1/2}M^{-1}$ is

\begin{equation}
f(r) = AM(r/\delta)^{2 \over n+2}\;,
\label{eq=fgb}
\end{equation}
where $A=(n+2)^{2/(n+2)}(n+4)^{-1/(n+2)} \sim 1$,
and $\delta = \lambda^{-1/2}M^{-1}$  is the size of the string core.
The energy density around the string is
$T_0^0 \propto r^{-2n/(n+2)}$
and the energy per unit length of string is

\begin{equation}
\mu (R) = 2\pi \int_0^R T_0^0 rdr \sim M^2 \left(R/\delta\right)^{4 \over n+2}
	\sim [f(R)]^2\;.
\label{eq=mugb}
\end{equation}
The cutoff radius $R$ has the meaning of a distance to the nearest 
string (or of the loop radius in the case of a closed loop).
We see from Eq.~(\ref{eq=mugb}) that vacuumless strings are very
diffuse objects with most of the energy distributed at large distances
from the string core.
They are much more diffuse than ordinary global strings which have
$\mu (R) \propto \ln R$, so that most of the energy is concentrated
near the core.

For the exponential potential~(\ref{eq=Vwallcosh}), 
numerical integration of Eq.~(\ref{eq=feqgb}) indicates that
the derivative terms in that equation are negligible at large $r$,
while the last two terms are nearly equal.
The asymptotic behavior of $f(r)$ at large $r$ is given by

\begin{equation}
f(r) \approx M\ln (r/\delta)
\label{eq=fcosh}
\end{equation}
and the dominant contribution to $\mu (R)$ is

\begin{equation}
\mu (R) \approx \int_\delta^R {f^2 \over 2r^2}2\pi rdr
	\approx {\pi \over 3} M^2[\ln(R/\delta)]^3\;.
\end{equation}

A global monopole is described by a triplet of scalar fields
$\phi_a$, $a=1,2,3$.
The monopole ansatz is $\phi_a=f(r)x_a/r$, where $r$ is the distance 
from the monopole center.
For the power-law potential~(\ref{eq=Vwallpower}),
it is easily verified that the field equation for $f(r)$ admits
a solution of the same form~(\ref{eq=fgb}) as for a global string
(with a different coefficient $A \sim 1$). The total energy of 
a global monopole is

\begin{equation}
{\cal M} \sim M^2R(R/\delta)^{4 \over n+2} \sim R[f(R)]^2\;,
\end{equation}
where the cutoff radius $R$ is set by the distance to the nearest antimonopole.
For the exponential potential~(\ref{eq=Vwallcosh}), $f(r)$ is
given by Eq.~(\ref{eq=fcosh}) and
${\cal M}(R) \sim 4\pi M^2R[\ln(R/\delta)]^2$.

\begin{figure}
\psfig{file=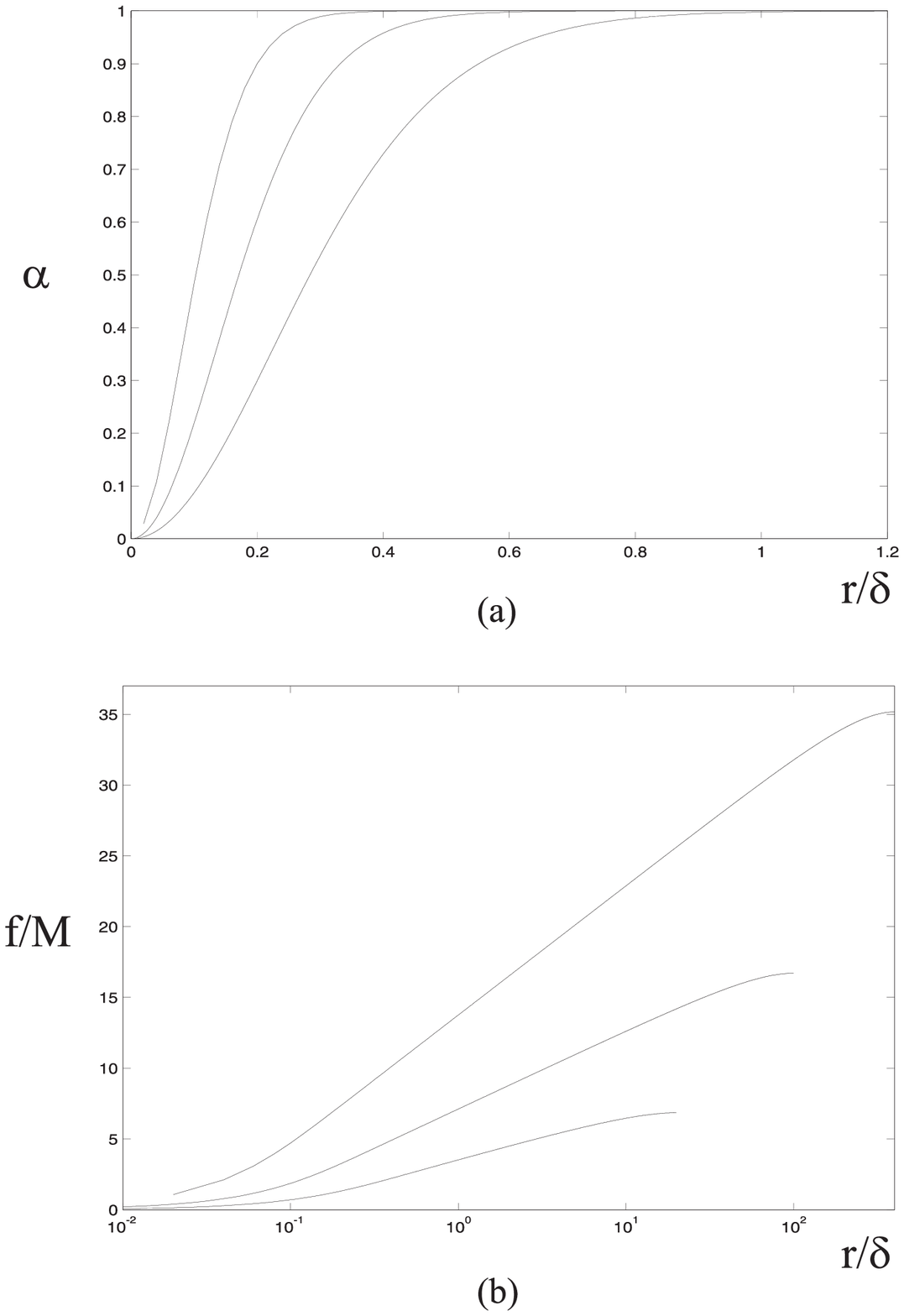}
\caption{
(a) Gauge field $\alpha (r)$ and
(b) scalar field $f(r)$ of a gauge vacuumless string 
for a potential~(\ref{eq=Vwallpower}) with $n=2$ and $e^2/\lambda =8$.
The top, middle, and bottom curves are for $R=400$, $100$, and $20$,
respectively.
}
\label{fig=nonvac}
\end{figure}

The structure of gauge vacuumless strings and monopoles is more unusual. 
A gauge symmetry breaking model is obtained from~(\ref{eq=lag}) by adding the
gauge field Lagrangian and replacing $\partial_\mu$ by gauge 
covariant derivatives.
We shall first consider strings with a power-law 
potential~(\ref{eq=Vwallpower}).
The string ansatz for the gauge field is $A_\theta (r)=-\alpha (r)/er$,
where $e$ is the gauge coupling. We found numerical  solutions to the field 
equations for $\alpha(r)$ and $f(r)$

\begin{eqnarray}
f''+{f' \over r} -(\alpha -1)^2{f \over r^2} -{dV \over df}=0
\label{eq=gsseq}\\
\alpha'' -{\alpha ' \over r} -e^2f^2(\alpha -1)=0
\label{eq=gsveq}
\end{eqnarray}
in a finite range $0 \leq r \leq R$,
with boundary conditions $f(0)=\alpha (0) = 0$, $\alpha (R) =1$.
The value of $f(R)$ was chosen so that the total field energy in the volume
under consideration is minimized. 
As $f(r)$ grows towards large $r$, the effective mass of the gauge field also grows, 
and $\alpha (r)$ approaches its asymptotic value, $\alpha =1$,
very quickly (see Fig.~\ref{fig=nonvac}a). 
The gauge flux tube is therefore very well localized and
the boundary condition $\alpha (R) =1$ is a good approximation.

The solution for $f(r)$ in the case $n=2$ and $e^2/\lambda =8$
is shown in Fig.~\ref{fig=nonvac}b for several values of $R$.
The most remarkable feature of these solutions is that, at a given $r$,
they do not approach any limit as the integration range $R$
is increased. This can be understood as follows. Outside the gauge flux tube,
we can set $\alpha =1$ and drop the third term in Eq.~(\ref{eq=gsseq}).
Then there are two possibilities. Either all three remaining terms 
in~(\ref{eq=gsseq}) are of comparable magnitude in the asymptotic 
region of large $r$, or two of the three are comparable, while the 
third term is negligible. Now, if the potential term is comparable
to one or both of the derivative terms, then the solution is 
a power law, Eq.~(\ref{eq=fgb}).
It is easily seen that the coefficient $A$ for this solution
cannot be a positive real number, and thus the solution is unphysical.
The only remaining possibility is that the two derivative terms are comparable
while the potential term is negligible.
This gives the solution

\begin{equation}
f(r)=A\ln (r/\delta) +B\;.
\label{eq=gssol}
\end{equation}
The problem with this solution is that with $f(r)$ given by~(\ref{eq=gssol}),
the potential term in Eq.~(\ref{eq=gsseq}) decreases more slowly than
the derivative terms. This indicates that~(\ref{eq=gssol}) cannot apply
in the whole asymptotic range $r \gg \delta$.
The potential term catches up with the derivative terms near the point
$r=R$, where the boundary conditions are imposed. 
Assuming that $B \lesssim A\ln (R/\delta)$ and requiring that
$f'' \sim dV/df$ at $r\sim R$, we have

\begin{equation}
A \sim M(R/\delta)^{2 \over n+2}[\ln (R/\delta)]^{-{n+1 \over n+2}}\;.
\label{eq=Ags}
\end{equation}
We have integrated Eqs.~(\ref{eq=gsseq}),~(\ref{eq=gsveq}) with $n=2$
numerically for several values of parameters $R/\delta$ and $\lambda /e^2$.
The results are in a good agreement with Eqs.~(\ref{eq=gssol}),~(\ref{eq=Ags}).

The energy per unit length of a gauge vacuumless string 
is $\mu (R) \sim 2\pi A^2(R)\ln (R/\delta)$, with the main
contribution coming from $f'^2/2$ and $V(f)$ terms in $T_0^0$.
Apart from a logarithmic factor, this is the same as Eq.~(\ref{eq=mugb})
for a global vacuumless string.
In the case of an exponential potential~(\ref{eq=Vwallcosh}),
a similar analysis yields $A\approx M$ and
$\mu (R) \sim \pi M^2 \ln(R/\delta)$.

The force per unit length due to inter-string interaction is
$F_i \sim d\mu (R)/dR$, and the force due to the string tension is
$F_t \sim \mu (R)/R$ (assuming that the typical curvature radius
of strings in the network is comparable to the average separation $R$).
For vacuumless strings with an exponential potential,
$F_t/F_i \sim \ln (R/\delta)$, and we expect the Nambu action
to give an adequate description of macroscopic strings 
($\ln (R/\delta) \sim 100$). For a power-law potential,
$F_i \sim F_t$, and the Nambu action cannot be used.

Finally, we consider vacuumless gauge monopoles. Once again, we find 
numerically that the potential term in the equation for $f(r)$
is negligible for $r \ll R$.
An analytic solution for both scalar and gauge fields can be
given in this range.
It is the Prasad-Sommerfield solution~\cite{Prasad}

\begin{eqnarray}
f(r) &=& A\coth (eAr) -1/er\;,\nonumber\\
\alpha (r) &=& eAr/\sinh (eAr)\;,\nonumber
\end{eqnarray}
where $\alpha (r)$ is defined in terms of the gauge field $A^a_\mu$
as $A^a_i = -[1-\alpha (r)]\epsilon^{aij}x^j/er^2$.
As before, $A$ can be estimated by requiring that the potential
term catches up with the gradient terms at $r\sim R$.
For the power-law potential, this gives

\begin{equation}
A(R)\sim M (\lambda eM^3R^3)^{1/(n+1)}\;.
\label{eq=Aggm}
\end{equation}
The size of the monopole core is $r_A \sim 1/eA$.
Much of the monopole energy is concentrated near the core;
the total energy in that region is

\begin{equation}
{\cal M} = {4\pi \over e}A\;.
\label{eq=monmass}
\end{equation}
In addition, there is a nearly constant energy density 
$T_0^0 \approx V(A)$ outside the core. The corresponding total energy
is ${\cal E}\sim V(A)R^3 \sim {\cal M}$. 
We note that the monopole core becomes a black hole
for $f(R) \approx A \gtrsim m_p$, where $m_p$ is the
Plank mass.
Quite similarly, for the exponential potential we find
$A \sim M\ln (R/\delta)$, while ${\cal M}$ is still given by 
Eq.~(\ref{eq=monmass}).

Let us now briefly discuss the formation and cosmological evolution
of vacuumless defects.
At high temperatures in the early universe, the potential $V(f)$
acquires the usual temperature-dependent term,

\begin{equation}
V_T(f) = aT^2f^2\;,
\label{eq=VT}
\end{equation}
where $a$ depends on the coupling of $\phi$ to other fields.
When $T$ is sufficiently large, the minimum of the potential is at $f=0$.
As $T$ decreases, this minimum turns into a maximum at some critical
temperature $T_c$.
For the exponential potential~(\ref{eq=Vwallcosh})
and for the power-law 
potential~(\ref{eq=Vwallpower}) with $n=2$,
$T_c=(\lambda/a)^{1/2}M$~\cite{TC}.
When the universe cools down to $T=T_c$, the symmetry-breaking phase 
transition occurs and the defects are formed in the usual way.

It should be noted that Eq.~(\ref{eq=VT})  for the thermal correction
to the potential is valid only at small $f$.
As $f$ gets large, particles coupled to $\phi$ acquire large masses
and disappear from the thermal bath.
As a result, $V_T(f)$ decreases exponentially at large $f$.
The thermal defect production mechanism will work only if $f$ is
initially localized near $f=0$.

An alternative way of triggering the symmetry-breaking phase transition
is to use the coupling of $\phi$ to the scalar curvature,
$V_{\cal R}(f)=\xi{\cal R}f^2$. The curvature ${\cal R}$ drops from
a large value to near zero at the end of inflation, and defect 
formation can occur at that point.
The phase transition can also be triggered by a coupling of $f$ to the
inflaton field.

By analogy with ordinary defects, one can expect that vauumless 
strings and monopoles will eventually reach a scaling regime in which
the typical distance between the defects is comparable to the
horizon, $R\sim t$. 
The main dissipation mechanism of vacuumless defects is similar
to that of ordinary global strings and monopoles
which lose most of their energy by radiating massless Goldstone bosons.
Global vacuumless defects will also radiate Goldstone bosons,
but in addition they will radiate massless $f$-particles
(which correspond to radial excitations of the field $\phi$).
For gauge vacuumless defects, there are no Goldstone bosons, but
$f$-particles will still be emitted.
Thus in all cases the dominant dissipation mechanism is
the radiation of massless particles.

For global strings with a power-law potential,
the mass per unit length of string is given by Eq.~(\ref{eq=mugb})
with $R$ replaced by $t$. The relative contribution of strings
to the energy density of the universe is given by

\begin{equation}
\rho_s/\rho \sim \mu (t)/m_p^2 \sim [f(t)/m_p]^2\;,
\end{equation}
where $f(t) \sim M(t/\delta)^{2/(n+2)}$ is
the characteristic magnitude of $\phi$ in the space between the strings.
We see that the fraction of energy in strings monotonically grows
with time, and the universe becomes dominated by the strings when
$f(t) \sim m_p$.

The observed isotropy of the cosmic microwave background (CMB) implies
$\mu(t_0)/m_p^2 \lesssim 10^{-5}$, where $t_0$ is the present time.
The corresponding constraint on $M$ is $M \lesssim 1$ MeV for $n=2$
and $M \lesssim 100$ GeV for $n=4$ (assuming $\lambda \sim 1$).
For values of $M$ saturating this constant, the strings would make a
non-negligible contribution to the observed CMB anisotropy on the largest
scales. Such strings, however, cannot be used to explain  structure formation.
The characteristic scale of the observed large-scale structure crossed 
the horizon at $t\sim t_{eq}\sim 10^{-6}t_0$. The density fluctuations
due to strings on that scale are of the order
$\delta\rho /\rho \sim \mu(t_{eq})/m_p^2 \ll 10^{-5}$.
Very similar conclusions are reached for gauge strings and global 
monopoles.

In the case of an exponential potential,
the evolution of global and gauge vacuumless strings is
similar to that of ordinary global strings, and the evolution of global
monopoles is similar to that of ordinary global monopoles.
The strings serve as seeds for structure formation
with $M\sim 10^{13}$ GeV and $M\sim 10^{15}$ GeV in the
global and gauge cases, respectively.
Global monopoles can seed structure formation for
$M \sim 10^{14}$ GeV.
The difference in the required energy scale $M$ for different
defects is due to the different power of the large 
logarithm in the expression for their mass.

Let us finally discuss
the evolution of vacuumless gauge monopoles.
A monopole and an antimonopole separated by a distance $R$
are attracted with a force $F\sim d{\cal M}/dR \sim {\cal M}/R$
(for the power-law potential)
and develop an acceleration $a\sim R^{-1}$.
They reach relativistic speeds and oscillate with a characteristic
period $\sim R$, rapidly losing their energy by radiation of massless
$\phi$-particles. Eventually they annihilate into $\phi$- and gauge
bosons. If $R$ is large enough, so that $f(R) > m_p$, then the 
monopole cores are black holes, and a Schwarzschild black hole of mass
$\sim {\cal M}$ (without a magnetic charge) is produced as a result of
annihilation.

Apart from close monopole-antimonopole pairs which are about to
annihilate, the typical distance between the monopoles at time $t$
is $R\sim t$, and their relative contribution to the energy
density of the universe is

\begin{equation}
\rho_m/\rho \sim {\cal M}(t)/m_p^2t \propto t^{2-n \over n+1}\;.
\end{equation}
For $n=2$, $\rho_m/\rho \sim (\lambda/e^2)^{1/3}(M/m_p)^2 = const$,
while for $n>2$ it decreases with time.

Monopoles turn into black holes at time $t_*$ such that $f(t_*)\sim m_p$.
The mass of black holes formed at time $t_f$ is $m\sim {\cal M}(t_f)$.
For $t_f$ in the radiation era, $t_f < t_{eq}$, the mass distribution
of black holes at the present time is

\begin{equation}
m{d\Omega_{BH} \over dm} \sim {mt_{eq}^{1/2} \over m_p^2t_f^{3/2}}
\sim 10^{27} \left({\lambda \over e}\right)^{1/2}
\left({M \over m_p}\right)^{5/2} 
\left({M \over em}\right)^{n-1 \over 2}\;,
\end{equation}
and for $t_f >t_{eq}$,
\begin{equation}
m{d\Omega_{BH} \over dm} \sim {m \over m_p^2t_f}
\sim  \left({\lambda \over e^2}\right)^{1/3}
\left({M \over m_p}\right)^2 
\left({M \over em}\right)^{n-2 \over 3}\;.
\end{equation}
The usual bounds~\cite{Carr} on the density of primordial black holes 
can be used to constrain the parameters $M$, $\lambda$, and $n$.

\acknowledgements
We are grateful to Allen Everett for useful comments
on the manuscript. This work was supported in part by 
the National Science Foundation.

\begin{references}
\bibitem{TC}
For $n>2$, the finite temperature 
potential always has a local minimum at $f=0$.
However, this minimum is protected by a potential barrier whose width and
height rapidly decrease with the temperature.
This barrier can be easily overcome by thermal fluctuations.
\end{references}
\end{document}